\documentclass[jcp,aip,preprint,longbibliography,floatfix,superscriptaddress]{revtex4-1}

\usepackage{bm}
\usepackage{afterpage}
\usepackage{amsmath,amssymb,amstext}
\usepackage{tabularx}
\usepackage{graphicx}
\usepackage{geometry}
\usepackage{float}
\usepackage{array}
\usepackage[export]{adjustbox}
\usepackage[para,online,flushleft]{threeparttable}
\usepackage[bookmarks=true,colorlinks=true,urlcolor=blue,linkcolor=blue,citecolor=blue]{hyperref} 
\newcommand{\temp}{k_{\text{B}}T}
\newcommand{\Ecut}{E_{\mathit{cut}}}
\newcolumntype{P}[1]{>{\centering\arraybackslash}p{#1}}
\newcolumntype{M}[1]{>{\centering\arraybackslash}m{#1}}

\newcommand*{\citen}[1]{%
  \begingroup
    \romannumeral-`\x 
    \setcitestyle{numbers}%
    \cite{#1}%
  \endgroup   
}

\begin{document}

  \title{\LARGE{Parameterization of magnetic vector potentials and fields for efficient multislice calculations of elastic electron scattering}}
  \author{\textbf{Keenan Lyon}}
\affiliation{\textnormal{$ ^1$\textsf{Department of Physics and Astronomy, Uppsala University, Box 516, S-751 20 Uppsala, Sweden}}}

\author{\textbf{Jan Rusz}}
\email{jan.rusz@physics.uu.se}
\affiliation{\textnormal{$ ^1$\textsf{Department of Physics and Astronomy, Uppsala University, Box 516, S-751 20 Uppsala, Sweden}}}

\date{\today}

\begin{abstract}

\noindent{\rule{0.91\textwidth}{1pt}}

\noindent{\bf\textsf{ABSTRACT}}

\noindent 
The multislice method \cite{cowley_scattering_1957}, which simulates the propagation of the incident electron wavefunction through a crystal, is a well-established method for analyzing the multiple scattering effects that an electron beam may undergo. The inclusion of magnetic effects into this method \cite{rother_relativistic_2009,edstrom_vortex_prl_2016} proves crucial towards simulating magnetic differential phase contrast images at atomic resolution \cite{edstrom_dpc_2019}, enhanced magnetic interaction of vortex beams with magnetic materials \cite{edstrom_vortex_prl_2016,edstrom_vortex_prb_2016}, calculating magnetic Bragg spots \cite{loudon_antiferromagnetic_2012}, or searching for magnon signatures \cite{lyon_2021_magnons}, to name a few examples. Inclusion of magnetism
poses novel challenges to the efficiency of the multislice method for larger systems, especially regarding the consistent computation of magnetic vector potentials \textbf{A} and magnetic fields \textbf{B} over large supercells. We present in this work a tabulation of parameterized magnetic values for the first three rows of transition metal elements computed from atomic density functional theory calculations \cite{GPAW, GPAWRev}, allowing for the efficient computation of approximate \textbf{A} and \textbf{B} across large crystals using only structural and magnetic moment size and direction information. Ferromagnetic bcc Fe and tetragonal FePt are chosen as examples in this work to showcase the performance of the parameterization versus directly obtaining \textbf{A} and \textbf{B} from the unit cell spin density by density functional theory calculations \cite{edstrom_vortex_prb_2016}, both for the quantities themselves and the resulting magnetic signal from their respective use in multislice calculations.

\noindent{\rule{0.91\textwidth}{1pt}}

\end{abstract}

\maketitle

\section{Introduction}

The engineering, design, and exploration of novel magnetic materials necessitates characterization methods capable of rendering the behaviour of these materials down to the atomic scale. Recent progress in the development of electron beam
monochromators has made it possible for the ever-versatile transmission electron microscope (TEM) to probe low energy excitations at this scale. Detection of magnetism in samples remains challenging, given that the interaction of magnetic moments with the electron beam is weaker than the Coulomb interaction by 3 to 4 orders of magnitude\cite{chapman_1978,rother_relativistic_2009,loudon_antiferromagnetic_2012}. Within the TEM setup, electron holography, Lorentz microscopy, differential phase contrast microscopy, and electron magnetic circular dichroism have all been put forward as approaches to study magnetism in materials. As these approaches gain momentum in the literature \cite{mcvitie_2015, Schattschneider_2006}, there is a clear need for a consistent and efficient description of magnetic vector potentials and fields in the materials under consideration. Efficiency becomes key in simulations of crystalline systems used in electron microscopy, where crystals on a size scale beyond the reach of standard DFT or other commensurate methods limits the computational reach for describing magnetism from \textit{ab initio} methods directly.

When it comes to a parameterization of a potential in the context of the multislice method, electron atomic scattering factors\cite{doyle_1968,weickenmeier_1991,peng_1999,peng_2005,kirkland_advanced_2010,lobato_2014}, first introduced to describe and evaluate the scattered beam amplitudes of electrons by crystals, come to mind. The use of the electron atomic scattering factors in this context relies on two main assumptions, namely that incoming electrons travelling at high enough energies will see the atom as a scattering center, and that the total Coulomb potential can be computed as a superposition of atomic potentials, neglecting the charge redistribution that occurs in a crystal lattice. While not identical, if similar criteria are assumed to hold true for magnetic fields and vector potentials with certain limitations, it stands to reason that knowledge of these quantities for an atomic setup can be used in superposition to build up a suitable approximation for the magnetic profile of any material.

For such a parameterization to be generally useful across a large spectrum of systems of interest in electron microscopy, certain criteria must be met. First, the difference in the atomic magnetic moment for the same atom in different crystal configurations must be easy to account for. Second, the potentials and fields should be smoothly varying independent of the choice of grid the quantities are represented on. Third, for computational efficiency, the determination of magnetic quantities at a given grid point should depend solely on local structural and magnetic moment size and direction information. This paper presents the quasi-dipole approach, satisfying all of the above criteria and thereby streamlining the implementation of magnetic potentials and fields, along with the associated Pauli multislice method\cite{edstrom_vortex_prl_2016,edstrom_vortex_prb_2016}, into the growing ecosystem of methods in microscopy that take account of magnetic effects in materials\cite{Rusz2018, edstrom_dpc_2019, lyon_2021_magnons, krizek2020atomically, Kovacs2017, Schneider2018, Midgley2009, Grillo2017, Matsumoto2016, Chen2018, Nguyen2020, Verbeeck2010}.

The methodology behind the parameterized magnetism method is outlined in Section~\ref{sec:method}, including the calculation of the periodic components of \textbf{A} and \textbf{B} from the spin density in DFT, the quasi-dipole approximation that forms the basis for parameterization of atomic magnetic components, and the computational details of the DFT calculations and the least-squares fit. Tabulated results for 25 transition metals are given in Section~\ref{sec:tabulate}, while the quality of the parameterized magnetism approach is benchmarked against DFT results and against different grids and geometries in Section~\ref{sec:eval}. Section~\ref{sec:multi} concludes the paper by showing how the parameterized magnetism approach utilized in the Pauli multislice method yields a magnetic signal that qualitatively matches with DFT results. 

\section{Methodology}\label{sec:method}

In the following sections we will summarize the methods used in this work. This begins in Section~\ref{sec:calc} with a summary of how the magnetic vector potential $\textbf{A}(\bm{r})$ and magnetic field $\textbf{B}(\bm{r})$ can be calculated using density functional theory (DFT) in a consistent manner, as shown in \cite{rother_relativistic_2009, edstrom_vortex_prl_2016}. The framework for the parameterization of fields surrounding atoms by fitting to DFT calculations is then developed in Section~\ref{sec:qda}. Computational details are summarized in Section~\ref{sec:comp}.

\subsection{Calculation of A and B fields}\label{sec:calc}
\noindent
The magnetization density for a crystalline system is given by
\begin{equation}
\bm{m}(\bm{r}) = \mu_B\langle \bm{\sigma} \rangle = \mu_B \mathrm{Tr}\left[\rho(\bm{r}) \bm{\sigma}\right] = \mu_B (2\textrm{Re} \left(\psi_\uparrow^* \psi_\downarrow \right), -2\textrm{Im} \left(\psi_\downarrow^* \psi_\uparrow\right), \rho_{\mathrm{spin}})
\end{equation}
where $\rho_{\mathrm{spin}} = \left|\psi_\uparrow\right|^2 - \left|\psi_\downarrow\right|^2$ is the spin density projected onto the spin quantization axis \cite{edstrom_vortex_prl_2016}. For the case of atomic systems, collinear magnetism occurs by default, resulting in the simplified
\begin{equation}\label{eq:matom}
\bm{m}_{\mathrm{atom}}(\bm{r}) = \mu_B\rho_{\mathrm{spin}}\hat{z}.
\end{equation}
In order to obtain the magnetic vector potential \textbf{A} and the corresponding flux density \textbf{B}, following Ref.~\citen{edstrom_vortex_prb_2016}, we first make the assumption that for the materials under consideration in this paper we can safely neglect the orbital current density. The total current density can therefore be expressed via \cite{strange_1998}
\begin{equation}
\bm{j} (\bm{r}) = \bm{j}_{\mathrm{spin}} (\bm{r}) = \nabla \times \bm{m}(\bm{r}).
\end{equation}
Working with Maxwell's equations in the Coulomb gauge ($\nabla \cdot \textbf{A} = 0$) yields
\begin{equation}\label{theory:deltaA}
\nabla \times \textbf{B}(\bm{r}) = - \Delta \textbf{A}(\bm{r}) = \mu_0 \bm{j}(\bm{r}).
\end{equation}
To solve for this given a magnetization density defined on a grid of a unit cell, we decompose the magnetic vector potential \textbf{A} into a periodic component $\textbf{A}_\mathrm{p}$, which corresponds to a zero average magnetic field $\textbf{B}_p$ in the unit cell, and a non-periodic component  $\textbf{A}_\mathrm{np}$ which in the Coulomb gauge is defined via
\begin{equation}\label{theory:np}
\textbf{A}_\mathrm{np} (\bm{r})= \frac{1}{2} \textbf{B}_\mathrm{avg} \times \bm{r}
= \frac{1}{2}(\mu_0 \textbf{M} + \mathbf{B}_\text{ext}) \times \bm{r},
\end{equation}
where \textbf{M} is the macroscopic magnetization of the material and $\mathbf{B}_\text{ext}$ is an external magnetic field. Since $\Delta \textbf{A}_\mathrm{np} (\bm{r})$ is here equal to zero, Eq.~\ref{theory:deltaA} can be written as
\begin{equation}
\Delta \textbf{A}_\mathrm{p}(\bm{r}) = -\mu_0 \bm{j}(\bm{r}).
\end{equation}
With periodicity in place, the magnetic vector potential and magnetic field can be expressed in reciprocal space as
\begin{equation}\label{theory:apbp}
\textbf{A}_\mathrm{p}(\bm{k}) = i\mu_0 \frac{\bm{k} \times \bm{m}(\bm{k})}{\bm{k}^2},\hspace{8mm} \textbf{B}_\mathrm{p}(\bm{k}) = -i \mu_0  \frac{\bm{k} \times (\bm{k} \times \bm{m}(\bm{k}))}{\bm{k}^2},
\end{equation}
where $\bm{m}(\bm{k})$ is the $\mathbf{k}$-component of the Fourier transformed atomic magnetization, see Eq.~\ref{eq:matom}. The average of $\textbf{B}_\mathrm{p}$ is zero due to the nature of the decomposition of the magnetic field, while the average of $\textbf{A}_\mathrm{p}$ in Eq.~\ref{theory:apbp} can be chosen to be zero by gauge freedom\cite{jackson_classical_1999}. For a DFT calculation where $\bm{m}(\bm{r})$ is defined on a grid, applying forward and backwards Fourier transforms will directly yield the periodic components of $\textbf{A}_\mathrm{p}$ and $\textbf{B}_\mathrm{p}$ in real space, with the zero average condition enforced by setting $\textbf{A}_\mathrm{p}(\bm{k}=\vec{0}) = \vec{0}$ and $\textbf{B}_\mathrm{p}(\bm{k}=\vec{0}) = \vec{0}$.

\subsection{Quasi-Dipole Approximation}\label{sec:qda}

In order to describe the magnetic vector potential and fields for each atomic system, we seek a function which can be parameterized in a way similar to electron form factors while still retaining the main properties of the dipole-like fields that surround atoms, a property that naturally comes about due to Hund's rule for the maximization of spin\cite{strange_1998}. For this approach we opt for a quasi-dipole formulation, such that the magnetic vector potential \textbf{A} and magnetic field $\textbf{B} = \nabla \times \textbf{A}$ are defined via
\begin{align}
&\hspace{7em}\textbf{A} = (\hat{m} \times \bm{r}) \sum_{i=0}^4 \frac{a_i}{r^{n(i)} + b_i}\label{theory:qda}\\
&\textbf{B} = \sum_{i=0}^4 a_i \frac{n(i)(\bm{r} \cdot \hat{m})r^{n(i)-2}\bm{r} + (2b_i - (n(i)-2))r^{n(i)}\hat{m} }{\left(r^{n(i)} + b_i\right)^2},\label{theory:qdb}
\end{align}
where $r = |\bm{r}|$, $n(i) = \frac{i}{2} + 3$, and $\hat{m}$ is the unit vector with direction of the magnetic moment of the atom. The $b_i$ coefficients in this equation serve to smooth out the short-range behaviour of the classical dipole, while the $a_i$ modulate the strength of the field over all space. It is important to note that the magnetic moment of an atom changes its magnitude depending on the surrounding environment\cite{billas_1994}, so considering materials with variable magnetic moments necessitates a rescaling of the magnetic moment vector. 

Utilizing the above equations serves three main purposes. First, introducing the parameters $b_i > 0$ eliminates the possible divergence associated with using a traditional dipole approximation. Second, by allowing for a sum over higher powers of the radial distance in the denominator, this approximation is better able to capture the short distance fluctuations within 1\AA~of the atomic centres that differentiate the magnetic properties of each element. Third, by hard-coding spherical symmetry i.e.~the shape of \textbf{A} and \textbf{B} will remain the same no matter which way the moment points, the parameterizations for each element can readily be used in the computation of larger structures, as only the element type and the direction and size of the magnetic moment are necessary to yield the periodic \textbf{A} and \textbf{B} components over all space. In addition, the symmetry of the fit function allows for the computation of radial prefactors that drastically speed up computation speed, namely
\begin{align}
\textbf{A} = &(\bm{\hat{m}} \times \bm{r}) a[r_\mathrm{q}] \hspace{1cm}\textbf{B} = \bm{r}(\bm{r} \cdot \bm{\hat{m}})b_1[r_\mathrm{q}] + \bm{\hat{m}} b_2[r_\mathrm{q}]\\
a[r] = \sum_{i=0}^4 \frac{a_i}{r^{n(i)} + b_i} &\hspace{0.5cm} b_1[r] = \sum_{i=0}^4 \frac{n(i)r^{n(i)-2}}{\left(r^{n(i)} + b_i\right)^2} \hspace{0.5cm} b_2[r] = \sum_{i=0}^4 \frac{(2b_i - (n(i)-2))r^{n(i)}}{\left(r^{n(i)} + b_i\right)^2}\nonumber
\end{align}
where $r_\mathrm{q}$ is a discretized approximation to $r$, depending on the grid spacing of the radial prefactors.

One major aspect where the atomic and bulk magnetic quantities diverge are in the quashing of the spin magnetic moment in the transition from atomic systems to clusters to the bulk \cite{billas_1994}, a natural consequence of the delocalization of atomic orbitals in response to bonding. This decreased local magnetic moment means that each individual atom will contribute less to the total \textbf{A} and \textbf{B} than in the atomic case. While a proper treatment of the shape of magnetic fields in response to orbital deformation needs to be considered on a material-specific basis, for the purposes of general approximation we present in this work that a simple scaling of the atomic parameterized fields to the experimental magnetic moment values is sufficient for the purposes of multislice calculations.  

In order for this parameterization to be generally useful towards the approximation of \textbf{A} and \textbf{B} for crystals of arbitrary size, the condition that the \textbf{A} and \textbf{B} within a certain area of each atom sums to zero is crucial, as this allows for the periodic component of \textbf{A} and \textbf{B} to be directly constructed from the individual zero-average atomic magnetic components. As seen in Eq.~\ref{theory:np}, the total magnetic field and vector potential for a system can then be obtained using the total magnetization. However, it is important to note that calculation of the parameterized values $a_i$ and $b_i$ in Eqs.~\ref{theory:qda} and \ref{theory:qdb} for atomic systems are optimized on a specific fine grid, so care must be taken that the final total sum of the periodic components of \textbf{A} and \textbf{B} over the entire supercell is as close to zero as possible.

\subsection{Computational Details}\label{sec:comp}

All DFT calculations employ the projector augmented wave method code \textsc{gpaw} \cite{GPAW,GPAWRev} within the atomic simulation environment \textsc{ase} \cite{ASE0,ASE}. An electronic temperature of $\temp \approx 1$~meV was chosen. All calculations are done in the spin polarized state. The Kohn-Sham wavefunctions are represented by plane waves (PWs) with a converged energy cutoff of $\Ecut \approx 400$~eV. For the 25 atomic calculations, Gamma point calculations were performed with a unit cell of dimensions $12\times12\times12$~\AA{}$^3$. The Perdew-Burke-Ernzerhof (PBE) parametrization \cite{PBE} of the generalized gradient approximation to the exchange-correlation (XC) functional was chosen for every element except for scandium, iron, nickel, rhodium, and osmium, for which the closely-related PBEsol\cite{PBEsol} XC functional was chosen due to convergence issues. The magnetic moments for scandium and palladium were forced into the $1\mu_B$ and $2\mu_B$ states respectively as otherwise these atomic systems proved difficult to converge. For bcc iron, defined by a $2.87 \times 2.87 \times 2.87$~\AA{}$^3$ unit cell with iron atoms located at $(0, 0, 0)$ and $(0.5, 0.5, 0.5)$ in scaled coordinates, and for FePt \cite{gilbert_2013}, defined by a $2.71 \times 2.71\times 3.72$~\AA{}$^3$ unit cell with iron at $(0.25, 0.25, 0.25)$ and platinum at $(0.75, 0.75, 0.75)$ in scaled coordinates, the PBE XC functional was again chosen while using a $6 \times 6 \times 6$ $k$-point Monkhorst-Pack mesh\cite{MonkhorstPack} for both calculations, noting that the relatively small $k$-point grid is sufficient for describing the approximate electron density in these lattices. The local magnetic moment calculated in \textsc{gpaw} for Fe in the bcc iron unit cell is $2.33\mu_B$, while in the FePt unit cell the local moment for Fe is $2.969\mu_B$ and for Pt is $0.397\mu_B$.

To determine the orientation of the magnetic moments for a supercell of bcc Fe of size~$30 \times 30 \times 100$ unit cells, angles $\theta$ for the magnetic moment divergence from the $z$-axis were sampled from a multivariate normal distribution of mean zero and standard deviation $30$\textdegree, simulating a supercell with $90\%$ of the $z$-direction magnetization of a collinear supercell, while the azimuthal angles $\phi$ were sampled uniformly from $0$\textdegree~to $360$\textdegree. An exponential distance decay factor\cite{Wackernagel_2003, rusz_2006} of $\exp\left(-\kappa d\right)$, where $\kappa=0.08$~\AA{}$^{-1}$ and $d$ is the distance between spins in \AA{}ngstr\"{o}m, was introduced into the covariance matrix for both distributions to imitate in-plane spin-spin spatial correlation, while spatial correlation along the $z$-axis was imitated by doing a layer-by-layer iterative mixing of $\theta$ and $\phi$, so every layer consists of a weighted average of $2/3$ the angles from the layer above and $1/3$ the angles drawn from the multivariate distributions.

\textbf{A} and \textbf{B} generated from the atomic DFT calculations yield grids of $108 \times 108 \times 108$ points over the $12\times12\times12$~\AA{}$^3$ cells. The parameterized values in Eqs.~\ref{theory:qda} and \ref{theory:qdb} are obtained with the LMFIT \cite{lmfit} package in Python, with optimization carried over the approximately 23000 points within a 2\AA{} radius of the atom centre and an additional 47000 points randomly chosen from elsewhere in the unit cell, with the only restriction on the least squares fit being that $b_i>0$ for all materials. A root mean squared error of at most 0.2 Tesla for the fits to \textbf{B} and 0.025 Tesla$\cdot$\AA{} for the fits to \textbf{A} were obtained.

\section{Results and Discussion}\label{sec:results}

In the following sections we present the results of this work. The tabulation of the parameterized magnetization values are first shown in Section~\ref{sec:tabulate}. Section~\ref{sec:eval} next shows a simulation for the appearance of the magnetic vector potential on a large supercell, the performance of the parameterized magnetism approach compared to DFT in describing magnetic fields in unit cells, and considers the flexibility of the approach with different grid sizes and geometries. Finally, Section~\ref{sec:multi} compares the magnetic signal from both a bcc iron and a tetragonal FePt supercell, using magnetic fields and potentials determined by the parameterized approach versus DFT as input to a Pauli multislice approach.

\subsection{Tabulated Magnetic Coefficients for the Transition Metals}\label{sec:tabulate}

\begin{table}
	\hspace*{-5em}
	\scalebox{0.75}{
		\begin{threeparttable}
			\hskip-8.0cm
			\begin{tabular}{|c|cc|cc|cc|cc|cc|cc|}
				\hline
				& \multicolumn{2}{c|}{Parameter 1}
				& \multicolumn{2}{c|}{Parameter 2}
				& \multicolumn{2}{c|}{Parameter 3}
				& \multicolumn{2}{c|}{Parameter 4}
				& \multicolumn{2}{c|}{Parameter 5}
				& \multicolumn{2}{c|}{RMS}\\\cline{2-13}
				Element & $a_0$ (T.\AA{}$^3$)& $b_0$ (\AA{}$^3$)& $a_1$ (T.\AA{}$^{3.5}$)& $b_1$ (\AA{}$^{3.5}$)& $a_2$ (T.\AA{}$^4$)& $b_2$ (\AA{}$^4$) & $a_3$ (T.\AA{}$^{4.5}$)& $b_3$ (\AA{}$^{4.5}$) & $a_4$ (T.\AA{}$^5$)& $b_4$ (\AA{}$^5$) & \textbf{A} (T.\AA{}) & \textbf{B} (T)\\
				\hline
				Sc & 9.627E-01 & 1.298E+00 & -1.195E-02 & 1.887E-03 & 3.207E-03 & 7.223E-04 & 1.746E-01 & 2.624E-01 & 2.062E-02 & 3.740E-02 & 2.70E-03 & 9.12E-03\\\cline{1-1}
                Ti & 9.865E-01 & 2.000E+00 & -2.565E-03 & 1.442E-03 & 3.906E-01 & 2.528E-01 & -5.238E-01 & 6.228E+00 & -1.731E-01 & 3.869E-01 & 1.32E-02 & 8.40E-02\\\cline{1-1}
                V & 6.594E+00 & 4.638E+02 & 2.007E+00 & 4.761E-01 & -4.102E+01 & 3.962E+03 & -1.165E+00 & 2.720E+00 & -7.793E-01 & 4.911E-01 & 1.37E-02 & 1.16E-02\\\cline{1-1}
                Cr & 1.201E+00 & 4.655E+00 & 1.304E+00 & 2.783E-01 & -5.867E+00 & 1.045E+00 & 3.577E+00 & 5.462E+00 & 3.880E+00 & 1.312E+00 & 2.15E-02 & 1.58E-02\\\cline{1-1}
                Mn & 8.182E-01 & 4.156E+00 & 2.390E+00 & 2.663E-01 & -6.858E+00 & 3.055E+00 & 6.222E+00 & 4.960E+00 & -7.540E-01 & 2.450E-01 & 1.35E-02 & 1.26E-02\\\cline{1-1}
                Fe & 6.967E+00 & 4.357E+02 & 1.799E+00 & 1.894E-01 & -4.485E+01 & 3.955E+03 & -6.665E-01 & 1.294E+00 & -5.191E-01 & 1.693E-01 & 1.27E-02 & 4.74E-02\\\cline{1-1}
                Co & 5.324E+00 & 2.716E+02 & 1.894E+00 & 1.638E-01 & -3.158E+01 & 2.386E+03 & -8.253E-01 & 9.567E-01 & -5.211E-01 & 1.285E-01 & 2.13E-02 & 1.80E-01\\\cline{1-1}
                Ni & 2.123E+00 & 2.109E-01 & 5.423E+00 & 6.535E+00 & -4.232E+01 & 9.541E-01 & 7.010E+01 & 1.139E+00 & -3.031E+01 & 1.325E+00 & 5.95E-03 & 7.90E-03\\\cline{1-1}
                Cu & 1.684E+00 & 1.976E-02 & -1.093E+00 & 5.604E-02 & -7.235E-02 & 2.058E-03 & -1.153E-01 & 4.499E-03 & -4.434E-01 & 5.774E-01 & 6.13E-03 & 3.65E-03\\\cline{1-1}
                Y & 1.808E+00 & 1.824E-03 & -5.816E-01 & 9.904E-02 & -1.824E+00 & 3.088E-04 & 1.045E+00 & 1.321E-04 & -1.714E-01 & 5.709E-05 & 6.13E-03 & 5.19E-02\\\cline{1-1}
                Zr & 2.954E-03 & 3.582E-04 & 1.251E+00 & 2.098E+01 & 9.615E-01 & 5.496E+00 & -1.008E-01 & 2.481E-01 & 1.946E-01 & 5.740E-01 & 1.45E-02 & 6.66E-02\\\cline{1-1}
                Nb & 9.850E-01 & 6.480E-02 & 3.354E-01 & 3.146E+00 & -3.161E-01 & 2.979E-02 & -1.071E+00 & 2.602E-01 & 6.619E-01 & 2.939E-01 & 1.79E-02 & 4.88E-02\\\cline{1-1}
                Mo & 1.150E+00 & 7.032E-01 & 2.098E-01 & 1.704E-02 & -1.331E+00 & 3.853E-02 & 7.077E-01 & 3.072E-02 & 1.815E-01 & 1.724E-01 & 1.75E-02 & 8.60E-02\\\cline{1-1}
                Ru & 1.141E+00 & 7.594E-01 & 2.508E-01 & 1.774E-02 & -1.329E+00 & 4.251E-02 & 5.850E-01 & 1.624E-01 & 3.849E-01 & 2.449E-02 & 1.56E-02 & 1.89E-01\\\cline{1-1}
                Rh & 2.283E-02 & 1.625E-03 & 2.769E+00 & 1.724E+00 & -2.187E+00 & 2.573E+00 & -1.444E-01 & 4.847E-02 & 1.402E-01 & 6.810E-02 & 1.60E-02 & 8.60E-02\\\cline{1-1}
                Pd & 7.623E-03 & 4.048E-04 & -3.558E+00 & 9.647E+01 & 9.942E+00 & 1.691E+02 & 2.203E+00 & 1.102E+01 & 2.391E-01 & 4.986E-01 & 6.11E-03 & 4.08E-02\\\cline{1-1}
                Ag & 4.349E+00 & 3.961E+01 & -8.312E+00 & 1.537E+02 & 4.611E-01 & 2.675E-01 & -6.189E-01 & 3.481E-01 & 2.320E+00 & 2.111E+01 & 2.01E-03 & 3.27E-03\\\cline{1-1}
                Hf & 8.602E-01 & 1.201E+00 & 2.697E-01 & 6.587E-02 & -1.946E-01 & 1.345E+00 & -1.715E-01 & 5.218E-02 & -1.149E-01 & 3.384E-01 & 9.93E-03 & 4.13E-02\\\cline{1-1}
                Ta & 9.782E-01 & 1.502E+00 & 2.160E-01 & 5.528E-02 & -6.757E-01 & 1.135E-01 & 3.900E-01 & 1.210E+00 & 2.566E-01 & 1.225E-01 & 1.26E-02 & 5.49E-02\\\cline{1-1}
                W & 8.705E-01 & 1.171E+00 & 9.543E-01 & 6.933E-02 & -1.546E+00 & 7.871E-02 & 2.710E-01 & 2.382E-01 & 3.197E-01 & 5.784E-02 & 1.71E-02 & 9.06E-02\\\cline{1-1}
                Re & 2.070E+00 & 6.697E+02 & 1.518E+00 & 9.766E-02 & 4.992E-01 & 1.397E+01 & -5.630E-01 & 4.738E-02 & -4.218E-01 & 2.385E-01 & 1.48E-02 & 9.59E-02\\\cline{1-1}
                Os & 1.042E+00 & 1.469E+00 & 3.499E-01 & 5.149E-02 & -1.514E+00 & 1.153E-01 & 1.086E+00 & 9.417E-01 & 7.026E-01 & 1.142E-01 & 1.23E-02 & 1.01E-01\\\cline{1-1}
                Ir & 2.514E-02 & 2.202E-02 & 1.139E+01 & 1.011E+02 & -2.154E+01 & 2.992E+02 & 2.050E+00 & 6.562E+00 & -2.924E-02 & 4.927E-02 & 1.45E-02 & 2.79E-02\\\cline{1-1}
                Pt & 2.001E+00 & 1.925E-01 & -3.305E+00 & 2.058E-01 & 1.185E+00 & 8.158E+00 & 3.716E+00 & 2.857E-01 & -2.088E+00 & 3.153E-01 & 5.79E-03 & 7.84E-02\\\cline{1-1}
                Au & 6.100E-01 & 1.747E+00 & 1.069E+00 & 2.511E+00 & 6.258E-01 & 2.764E-01 & -3.433E+00 & 6.340E-01 & 2.102E+00 & 7.760E-01 & 2.67E-03 & 1.16E-02\\\cline{1-1}
				\hline
			\end{tabular}
		\caption{\large Parameterized magnetic factors for transition metal elements from scandium ($Z = 21$) to gold ($Z = 79$), following Eqs.~\ref{theory:qda} and \ref{theory:qdb}. Elements with filled $d$-orbitals or for which \textsc{gpaw} has no atomic PAW setups are left out. The parameters $a_i$ are scaled so that the resulting \textbf{A} and \textbf{B} values correspond to a spin magnetic moment of one Bohr magneton. The root mean square error (RMS) is included for each calculation.}
		\end{threeparttable}
	}
	
\end{table}

Table~1 shows the parameterized magnetic factors for transition metal elements from scandium ($Z = 21$) to gold ($Z = 79$), following Eqs.~\ref{theory:qda} and \ref{theory:qdb}, neglecting elements with filled $d$-orbitals (and therefore zero spin magnetic moment) or for which \textsc{gpaw} has no atomic PAW setups (Tc). The parameters $a_i$ are scaled so that the resulting \textbf{A} and \textbf{B} values correspond to an atom with spin magnetic moment of one Bohr magneton. Therefore, for example, calculations for a bcc iron supercell would involve a rescaling of the listed $a_i$ parameters in Table~1 by a factor of $2.33$ for every constituent iron atom.

\begin{figure}
	\includegraphics[width=1\textwidth, center]{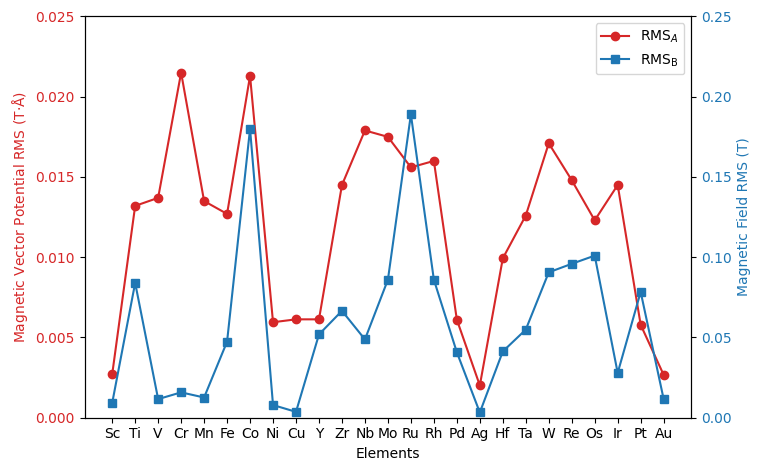}
	\caption{Root mean squared error (RMS) for least squares fit of parameterized values in Eqs.~\ref{theory:qda} and \ref{theory:qdb} versus atomic DFT calculated \textbf{A} and \textbf{B} values.
	}\label{fig:rms}
\end{figure}

Three main features stand out from Table~1. First is the fact that the values of $a_4$, corresponding to a quasi-dipole that propagates asymptotically in space as $1/r^5$ rather than the $1/r^3$ of the classical dipole\cite{jackson_classical_1999}, have a median an order of magnitude lower than values $a_0$ to $a_3$, suggesting that the fluctuations in the magnetic fields located very close to the atomic centres are not crucial to the overall performance of the model. Second is that all values $a_0$ are positive across all elements, which matches with the expectation that since these coefficients correspond to the quasi-dipole term closest to the classical dipole, the contribution to the overall magnetization is positive as well. Third, the median values for $b_i$ are between $0.25$ and $1.25$ in units of \AA{}$^{n(i)}$, suggesting again that on the whole no one term in the quasi-dipole approximation is accounting for short or long term behaviour of the \textbf{A} and \textbf{B} values over the entire unit cell. The root mean squared error for the calculations over the 25 transition metal elements is presented in Fig.~\ref{fig:rms}. A general relationship exists between parameterizations that yield good fits for \textbf{A} also doing so for \textbf{B}, but most importantly the maximum error of this approach is revealed to be consistent across a range of atomic elements.

\begin{figure}
    \hskip1.0cm
    \vskip-2.0cm
	\includegraphics[width=0.94\textwidth, center]{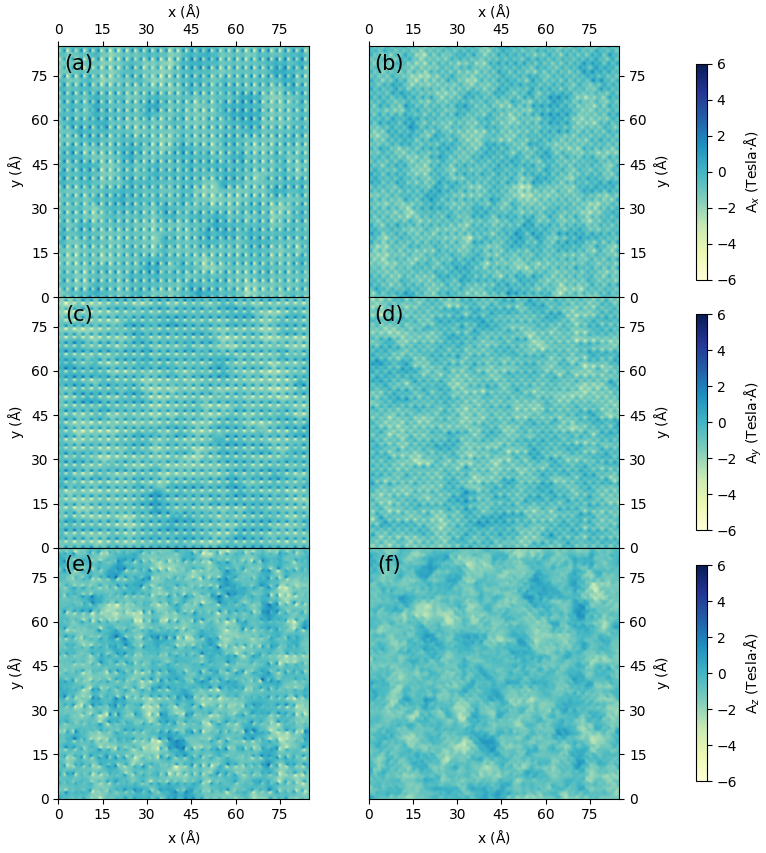}
	\caption{Heatmaps in the $xy$-plane for (a, b) $A_x$ (c, d) $A_y$ and (e, f) $A_z$ for a supercell of bcc Fe of size $86 \times 86 \times 287$~\AA{}$^3$, with magnetic vector potential generated from the parameterization of magnetic fields around the iron atom. Directions of the magnetic moments are given by spatially correlated multivariate normal and uniform distributions. (a,c,e) are for the plane located at $z=0$~\AA{}, where 900 atoms lie at the surface, while (b,d,f) are for the plane located at $z=0.75$~\AA{} deeper in the material.
	}\label{fig:mvp}
\end{figure}

\subsection{Evaluation of the Parameterized Magnetism Approach}
\label{sec:eval}

To showcase the capabilities of the parameterized magnetism (PM) approach, Fig.~\ref{fig:mvp} presents heatmap plots in the $xy$-plane for three components of \textbf{A} for a supercell of bcc Fe of size $86 \times 86 \times 287$~\AA{}$^3$ i.e.~$30 \times 30 \times 100$ unit cells. The orientation of the magnetic moments for this system were given by two spatially correlated multivariate distributions as explained in Section~\ref{sec:comp}. The magnetic field and vector potential were evaluated on a grid of $1500 \times 1500 \times 3000$ points. In Fig.~\ref{fig:mvp}, the subfigures in the left column are for the plane located at $z=0$~\AA{}, where 900 atoms lie at the surface. A sort of grid-like pattern emerges for $A_x$ and $A_y$, as nearly all moments are oriented towards the $z$-axis and the fluctuations in the spin density are strongest nearest to the atoms. Fewer of these punctures are visible for $A_z$ as the local moments there must point relatively off of the $z$-axis. A broad continuity of the colour spectrum is also visible, reflecting the slow fluctuations expected from the chosen spatial autocorrelation factor. The lack of sharp peaks or troughs in the vector potential is a reflection of the introduction of $b_i$ terms in Eqs.~\ref{theory:qda} and \ref{theory:qdb}, which would not be the case for certain choices of grid in a classical dipole approach.

The right column subfigures of Fig.~\ref{fig:mvp} are for the plane located at $z=0.75$~\AA{} deeper in the material. It is evident that the general fluctuations in the three components of \textbf{A} match with those of the left column of subfigures. The smoother nature of these heatmaps versus the left column is a natural consequence of being in a plane an equal distance from both planes of iron atoms.

\begin{figure}[htp]
    \vskip-1.0cm
	\includegraphics[width=0.98\textwidth, center]{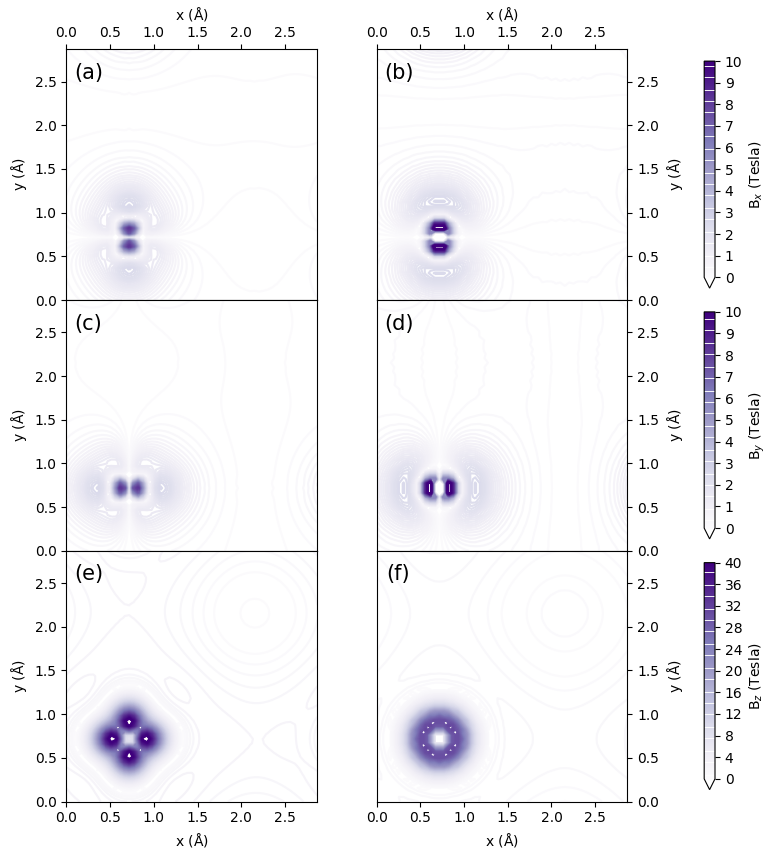}
	\caption{Contour plots in the $xy$-plane of a bcc iron unit cell for (a, b) $B_x$ (c, d) $B_y$ (e, f) $B_z$ along the $z$-plane of one of the iron atoms, with (a, c, e) calculated directly from the DFT calculated spin density and (b, d, f) calculated via the parameterized values shown in Table~1 for the iron atom, with magnetic moments normalized to bulk values.}
	\label{fig:bcc}
\end{figure}

\begin{figure}[htp]
    \vskip-1.0cm
	\includegraphics[width=0.97\textwidth, center]{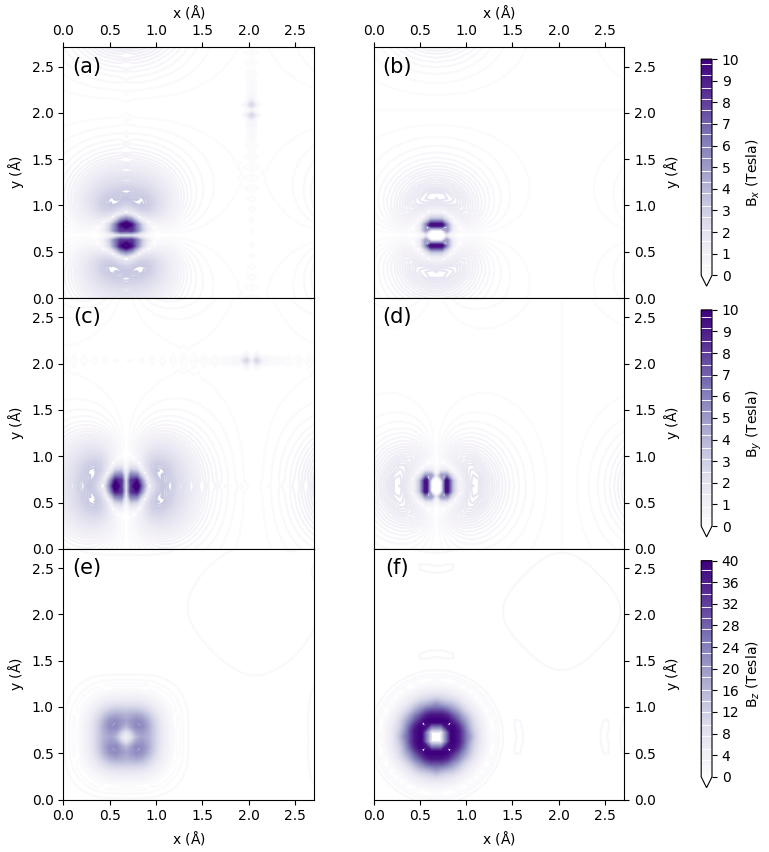}
	\caption{Contour plots in the $xy$-plane of a FePt unit cell for (a, b) $B_x$ (c, d) $B_y$ (e, f) $B_z$ along the $z$-plane of the iron atom, with (a, c, e) calculated directly from the DFT calculated spin density and (b, d, f) calculated via the parameterized values shown in Table~1 for the iron and platinum atoms, with magnetic moments normalized to bulk values.}
	\label{fig:fept}
\end{figure}

Aside from the performance of the parameterized magnetism approach in generating magnetic vector potentials and fields over large supercells, it is instructive to see the predictive capabilities of this approach by contrasting with \textbf{A} and \textbf{B} generated directly from the DFT supercell. Figs.~\ref{fig:bcc} and \ref{fig:fept} show contour plots of the three components of \textbf{B} in the $xy$-plane of a collinear setup for a periodic bcc iron unit cell and a periodic FePt unit cell, respectively. Both sets of subfigures evaluate the magnetic fields at the $z$-axis location of the topmost iron atom, with the left column showing results from the converged spin density of a DFT calculation over a unit cell and the right column showing the parameterized magnetization approach, including the scaling of \textbf{A} and \textbf{B} by the bulk moments as listed in Sec.~\ref{sec:comp}.

For Fig.~\ref{fig:bcc}, the approximation of the magnetic fields along the $x$- and $y$- directions are in close agreement both in magnitude and shape. Along the $z$-direction, the contours in Fig.~\ref{fig:bcc}(f) reveal the underlying symmetry inherent in the quasi-dipole approximation, as the shapes reflecting the more orbital-like electron distribution due to bonding as seen in Fig.~\ref{fig:bcc}(f) are not present in this figure. In both (e) and (f) the presence of the second iron atom in the top left can also be faintly seen. As the magnitudes along this direction are also in close agreement, it is likely that the parameterized magnetization approach will serve as a close approximation to the magnetic behaviour for this system.

For Fig.~\ref{fig:fept}, we again see qualitatively that the general shape along all directions for the \textbf{B} field are in good agreement, with the rightmost column showing parameterized magnetic fields having a more symmetrical character than their DFT counterparts. However, in contrast to bcc iron in Fig.~\ref{fig:bcc}, it is noticeable that along the $x$- and $y$- directions the PM approach underestimates the DFT field by a factor of $0.6$, while along the $z$-direction the \textbf{B} field is overestimated by a factor of $1.55$. For this material, deformation of the electron density surrounding the iron atoms in response to neighbouring platinum atoms has changed the surrounding magnetic vector potentials and fields in such a way that our quasi-dipole approximation, which enforces a fixed ratio between the three directions of each quantity, cannot provide a suitable quantitative fit along every direction. It is expected that if calculations like the multislice method are more dependent on a given direction of \textbf{B} for a particular setup, the results for the parameterized magnetization approach may provide a quantitatively worse description for the magnetic properties of the system. The consequences of this anisotropy of spin density will be evaluated in Sec.~\ref{sec:multi} below.

\begin{figure}[htp]
	\includegraphics[width=0.8\textwidth, center]{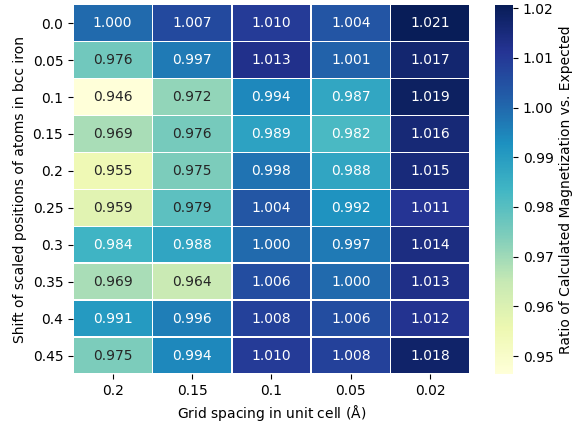}
	\caption{Heatmap of the ratio of the calculated total magnetization versus the expected for a collinear unit cell of bcc iron, using grid spacings ranging from $0.02$\AA{} to $0.2$\AA{} and with the locations of iron atoms located in the scaled coordinates range between $\langle (0,0,0), (0.5, 0.5, 0.5)\rangle$ (shift of $0$) and $\langle (0.45,0.45,0.45), (0.95, 0.95, 0.95)\rangle$ (shift of $0.45$).}
	\label{fig:grids}
\end{figure}

Returning to a point made in Sec.~\ref{sec:qda}, we explore numerical behaviour in utilizing the parameterized magnetism approach with regards to grid sizes and sparsity. For the first aspect, as the parameterization for each atom is calculated and fit on a $0.11$\AA{} spaced grid, it is expected that grids of different sizes, especially coarser ones, may have a substantial effect on the total magnetization within a cell. The second aspect regarding sparsity relates to the mismatch between unit cell lattice parameters and the grid spacing, as for sparser grids the fields around each atom will not be sampled as evenly or as symmetrically as for a fine grid. In Fig.~\ref{fig:grids}, calculations of \textbf{A} and \textbf{B} over $5 \times 5 \times 5$ supercells of collinear bcc iron were done with the parameterized magnetism approach, with the unit cells varying between those with atoms located at $\langle (0,0,0), (0.5, 0.5, 0.5)\rangle$ in scaled coordinates to those at $\langle (0.45,0.45,0.45), (0.95, 0.95, 0.95)\rangle$ in steps of $0.05$. Grids of size $0.02$\AA{}, $0.05$\AA{}, $0.1$\AA{}, $0.15$\AA{}, and $0.2$\AA{} were used for each system. The expected magnetization per unit cell is given by the two iron atoms each with local magnetic moment of $2.33\mu_B$.

Two main features stand out in Fig.~\ref{fig:grids}. First, for a grid spacing lower than $0.1$\AA{}, the total magnetization per unit cell of bcc iron will be within $\pm 2\%$ of the expected value, reflecting the smoothness of the quasi-dipole approximation and its portability to fine grids of different sizes. It is worth noting that the higher magnetization ratio for the finer grid is not necessarily the case across the parameterization of every element and should be considered on a case-by-case basis. Second, as expected, sparse grids have a strong influence both on the magnetization ratio relative to the finer grids and between grids of the same size but with geometrically isomorphic unit cells. Most importantly, the choice of a suitably fine grid for calculations of the magnetic vector potential and fields should yield a magnetization in line with calculations optimized on the atomic DFT grid.

\begin{figure}[htp]
	\includegraphics[width=1.1\textwidth, center]{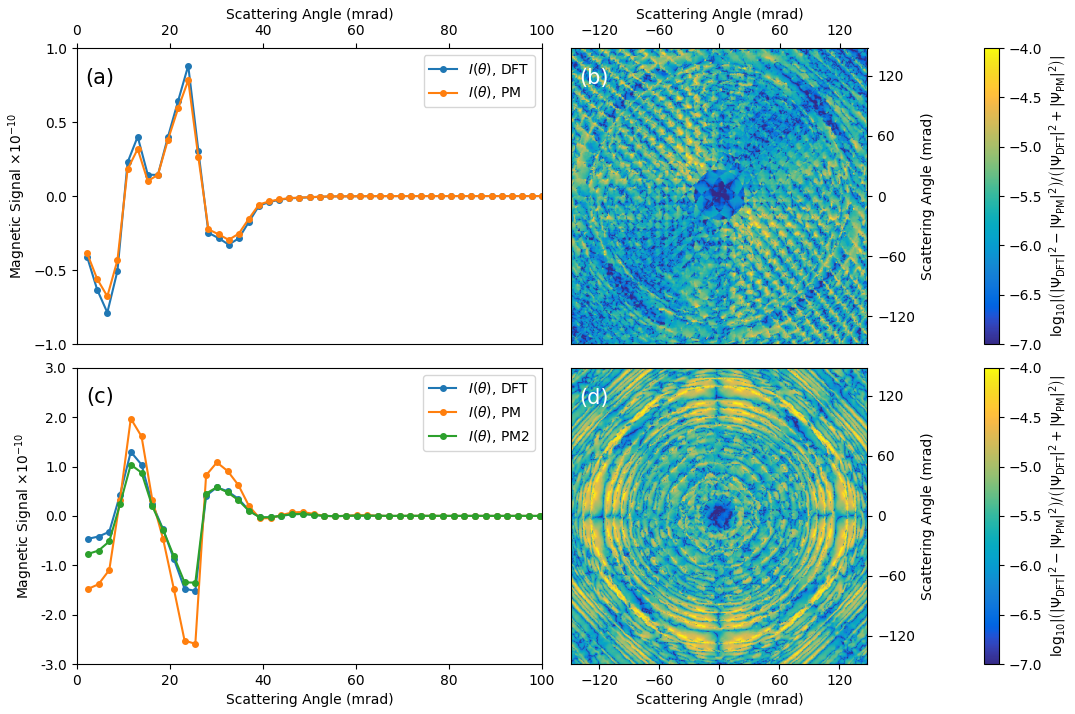}
	\caption{Magnetic signal for collinear (a) bcc iron and (c) FePt using magnetic vector potentials and fields calculated using DFT and PM respectively, with PM2 for the FePt showing the magnetic signal with PM parameters scaled directly using DFT fields instead of bulk magnetic moments. The logarithm of the relative ratio of the squared amplitude of the output wavefunctions for the DFT and PM methods are given for (b) bcc iron and (d) FePt. Output wavefunctions are calculated with the Pauli multislice method\cite{edstrom_vortex_prb_2016} using $20 \times 20 \times 10$ unit supercells with $V_{\mathrm{acc}} = 200kV$ and $\alpha = 25$~mrad.}
	\label{fig:multislice}
\end{figure}

\subsection{Magnetic Signal in the Multislice Method Using Parameterized Magnetism}\label{sec:multi}

Most importantly for the parameterized magnetism approach, from the perspective of performing multislice calculations, is its predictive capability for the magnetic signal for large supercells. For the case where the magnetic moments in a large supercell all point in different directions, doing DFT simulations to determine the magnetic field across the entire cell is challenging computationally. However, for a fully collinear system, the periodic magnetic vector potential and field in any one unit cell will be identical, while the non-periodic component can be computed via Eq.~\ref{theory:np}, and the results of multislice calculations from both DFT and parameterized magnetism can be directly compared.  

For Fig.~\ref{fig:multislice}, supercells consisting of $20 \times 20 \times 10$ unit cells for both bcc iron and FePt were considered. An acceleration voltage of $V_{\mathrm{acc}} = 200kV$ was used along with a convergence semi-angle of $\alpha = 25$~mrad. A magnetic field of $2$~T was added to both crystals. Multislice calculations were performed with zero periodic magnetic components (i.e.~all moments set to zero) for calibration, with the DFT calculated magnetic fields for bcc iron and FePt as outlined in Section~\ref{sec:comp}, and with parameterized magnetism (PM) calculations done using the values of bulk moments obtained from DFT as listed in Section~\ref{sec:comp}. In addition, for the FePt, calculations were also carried out with the tabulated values for iron normalized instead by $\mu_{\mathrm{Fe, local}} = 1.687\mu_B$ and $\mu_{\mathrm{Pt, local}} = 0.39\mu_B$. This is done in order to explore the effect of having the $B_z$ for the FePt system with moments aligned along the $z$-axis match quantitatively with the DFT results as seen in Fig.~\ref{fig:bcc}. This approach has been labelled as PM2 in Fig.~\ref{fig:multislice}.

Plotted in subfigures (a) and (c) of Fig.~\ref{fig:multislice} are the radial magnetic signal after subtracting off the squared amplitude of the calibration exit wavefunction, $\Psi_\mathrm{CAL}$, at each pixel from that of the calculated exit wavefunction, given by
\begin{equation}
I_{\mathrm{DFT/PM/PM2}} (\theta) = \frac{\sum_{\theta - \Delta \theta \leq r < \theta} \left[\left|\Psi_{\mathrm{DFT/PM/PM2}} (i,j) \right|^2 - \left|\Psi_{\mathrm{CAL}}(i,j) \right|^2\right]}{\sum_{\theta - \Delta \theta \leq r < \theta} 1},
\end{equation}
where $i, j$ are the pixel positions relative to the centre of the diffraction pattern, $r = \sqrt{i^2+j^2}$, ``CAL" refers to the calibrated exit wavefunction from only having the $2$~T field applied to the supercell, and a $\Delta \theta$ of $2$~mrad was chosen. Both subfigures reveal that the PM approach is to a strong degree able to qualitatively predict the magnetic signal in these large supercells to a similar degree as DFT generated magnetic vector potentials and fields across all areas of the diffraction pattern. For the bcc iron in (a), the PM consistently underestimates the DFT magnetic signal by a factor between $5$ and $10\%$, matching with the expectation hinted at in Fig.~\ref{fig:bcc} that the two magnetic quantities were qualitatively and quantitatively similar. For FePt in (c), the PM consistently overestimates the magnetic signal relative to DFT, with the smallest scattering angles especially showing a difference of the order of $200\%$ difference. For this reason the PM2 approach is included, recalibrating the magnetic moments so that the $B_z$ fields are in close quantitative agreement. The PM2 approach clearly works to bring the magnetic signal more in line with the DFT predicted value, suggesting that this may be a superior approach for systems with strong deformation of the electronic density around the concerned atoms. (b) and (d) in Fig.~\ref{fig:multislice} show the logarithm of the relative ratio (i.e.~difference divided by the sum) of the squared amplitude of the DFT and PM exit wavefunctions for the bcc iron and FePt supercells respectively, providing a visual clue as to the degree these two approaches are in agreement, with the bcc iron relative ratio consistently below $10^{-5}$ across the whole diffraction pattern while the FePt relative ratio appears most prominent in the region beyond $60$~mrad from the centre, which as seen in the magnetic signal plots in (a) and (c) is where the magnetic signal is already asymptotically trending towards zero.

\section{Conclusion}

A framework for the atomic parameterization of magnetic vector potentials and fields for transition metal elements has been presented herein, with the overarching goal being to provide an efficient and reliable method for the inclusion of magnetic effects in magnetic multislice calculations\cite{edstrom_vortex_prb_2016} for materials and crystals of arbitrary size. Calculating these magnetic quantities traditionally requires either a heavy effort on the part of computationally demanding software, or on locally inaccurate approximations like a classical dipole method\cite{jackson_classical_1999}. Relying on spin densities generated in \textsc{gpaw}\cite{GPAW, GPAWRev}, a quasi-dipole approximation consisting of 10 free parameters was fit using least squares for 25 transition metal elements\cite{lmfit}. The flexibility of this approach was showcased by a magnetic vector potential \textbf{A} on a grid of size $85 \times 85 \times 283$~\AA{}$^3$ for a bcc iron supercell with moments aligned according to spatially correlated normal and uniform distributions. The performance of the parameterized magnetization was directly compared with magnetic quantities derived from DFT calculations in the unit cell for bcc iron and for tetragonal FePt, showing that the performance of the parameterization is best for materials without significant deformation of their spin density due to bonding\cite{billas_1994}. The performance of the parameterized magnetization approach was shown to be flexible on grids and geometries of different sizes. Lastly, a direct comparison of the magnetic signals resulting from Pauli multislice calculations of the approach in contrast with DFT calculations showed that for both bcc iron and tetragonal FePt, the parameterized magnetism method was able to capture the behaviour of the magnetic signal as a function of scattering angle, with better quantitative results depending on the scaling of magnetic moments in the unit cell.

Computational work employed resources from Swedish National Infrastructure for Computing (SNIC) as well as from the Imbabura cluster of Yachay Tech University, which was purchased under contract No. 2017-024 (SIE-UITEY-007-2017).

\section{ORCID IDs}
\noindent
K Lyon \href{https://orcid.org/0000-0002-4374-2077}{https://orcid.org/0000-0002-4374-2077}\\
J Rusz \href{https://orcid.org/0000-0002-0074-1349}{https://orcid.org/0000-0002-0074-1349}

\bibliographystyle{apsrev4-1}
\bibliography{references}

\end{document}